# Modeling and Optimization of Transistor Voltage Amplifiers Based on Stochastic Thermodynamics

Xiaoxuan Peng,[1,2,3] Xiaohu Ge,[1,2,3*]

[1] School of Electronic Information and Communication, HuaZhong University of Science and Technology, Wuhan 430074, China

[2] International Joint Research Center of Green Communications and Networking, Wuhan 430074, China

[3] These authors contributed equally to this work.

* Correspondence author, e-mail address: xhge@mail.hust.edu.cn.

**ABSTRACT**

As transistor sizes reach the mesoscopic scale, the limitations of traditional methods in ensuring thermodynamic consistency have made power dissipation optimization in transistor amplifiers a critical challenge. Based on stochastic thermodynamics, a transistor voltage amplifier model is first proposed as a new insight to investigate nonlinear relationships between the power dissipation and voltage gain for complementary symmetric voltage amplifier circuits (CSVACs). Utilizing the proposed model, the phenomenon, i.e., the power dissipation exponentially increases with the increase of voltage gain in CSVACs, is first clarified by an analytical expression to quantify the impact of voltage gain and input signal amplitude on the power dissipation. Considering the characteristic of power dissipation, a new multistage architecture is proposed to reduce the power dissipation in CSVACs. To optimize the power dissipation by adjusting the number of stages and the voltage gain at each stage, an optimal multistage scheme is proposed for multistage CSVACs. Simulation and experimental results show up to 99.36% and 94.59% power dissipation reduction compared with traditional CSVACs by the proposed optimal multistage scheme, respectively.

**KEYWORDS**

**1 INTRODUCTION**

Signal amplification is widely applied across various domains, such as aiding in overcoming signal attenuation and noise in communication systems[1,2], facilitating the measurement and analysis of weak signals in diverse fields of fundamental research[3-7], and being employed for the diagnosis and monitoring of physiological processes in biomedical applications[8-10]. Based on the market predicted data released by Global Information, the global market size of amplifiers based on complementary metal-oxide-semiconductor (CMOS) was 5.69 billion dollars in 2025, which was predicted to reach 9.82 billion dollars by 2032[11]. However, the amplification process is often accompanied by substantial energy dissipation and efficiency loss due to the indiscriminate noise amplification[12] and incomplete conversion of input energy into output signal[13].

The power dissipation of signal amplification typically refers to the power dissipation during operations of signal amplifier circuits. Presently, signal amplifier circuits widely adopt the transistor amplifier effect to realize signal amplification. To achieve a low power dissipation of transistor signal amplifier circuits, some approaches, such as reducing the number of electrons[14], lowering the supply voltage[15], and reducing the transistor size by shortening lithographic processes[16], are primarily adopted. However, reducing the number of electrons and the supply voltage can also lead to the degraded performance of signal amplifier circuits, such as decreased gain and increased electron thermal noise within transistors[17]. Enhancing the intrinsic performance of transistors becomes increasingly challenging when transistor dimensions enter the mesoscopic scale (1-100 nanometers)[18,19]. When transistor sizes reach the mesoscopic scale, traditional analytical methods linearly treat the noise from a phenomenological perspective, failing to ensure the thermodynamic consistency between the environment and the transistor[20]. Consequently, traditional analytical methods cannot precisely analyze the non-equilibrium information processing of transistors when energy consumption levels approach the Landauer limit. Therefore, modeling and analyzing the non-equilibrium information processing of transistors are emerging as crucial and unignorable issues for transistor-based signal amplifier circuits when the transistor size approaches the mesoscopic scale.

Stochastic thermodynamics is commonly used to address dynamic system problems using stochastic processes. It provides a theoretical foundation for analyzing the non-equilibrium information processing processes of transistors at the mesoscopic scale[21]. Existing relevant studies predominantly focus on the construction and thermodynamic description of physical models for electronic devices such as tunnel junctions and diodes[20,22-24]. A few related studies extend to the analysis of specific circuits, including voltage and current fluctuations caused by thermal noise in logic circuits[25], the interplay among accuracy, speed, and energy consumption during logic gate operations[26], modeling and optimization of energy consumption during logic circuit operations, and the interaction between the information and energy[27-30]. However, existing studies predominantly focus on digital logic circuits, with little investigation conducted on the amplifier effects of transistors and transistor-based signal amplifier circuits.

Common signal amplification circuit structures include differential amplification structures, push-pull amplification structures, and complementary symmetry amplification structures. Among these, the signal amplifier circuit with complementary symmetry structure, i.e., the complementary symmetric voltage amplifier circuits (CSVAC), is widely used in various communication systems due to its high linearity, high thermal stability, and low noise characteristics[31,32]. Based on stochastic thermodynamics, a stochastic model of single-level transistors is proposed for describing the power dissipation of transistors with non-equilibrium information processing processes. The proposed model is able to describe the electron transfer process consistent with the fluctuation theorem while guaranteeing compatibility with the CMOS infrastructure[26]. Based on the single-level transistor stochastic model, a stochastic thermodynamics model is established for the CSVAC. Utilizing the stochastic thermodynamics model, the nonlinear relationship among the power dissipation, the input signal amplitude and voltage gain is first quantified in CSVACs. Moreover, experiment results validate the nonlinear relationship among the power dissipation, the input signal amplitude and voltage gain in CSVACs. Based on the nonlinear relationship among the power dissipation, the input signal amplitude and voltage gain, a multistage architecture is proposed to reduce the power dissipation of CSVACs' non-equilibrium information processing processes. Furthermore, an optimal multistage scheme is derived for multistage CSVACs, i.e., the optimal configuration of the number of stages and voltage gain at each stage. Compared with the power dissipation of traditional CSVACs, simulation and experiment results demonstrate that the power dissipation of multistage CSVACs can be reduced by up to 99.36% and 94.59% by the proposed optimal multistage scheme, respectively.

## 2 MATERIALS AND METHODS

### 2.1 Numerical simulation

The stochastic thermodynamic models of transistors, transistor voltage amplifier circuits, and CSVACs were all implemented in Python and executed on a Personal Computer (PC). The power dissipation model of multistage CSVACs was built with MATLAB to obtain the fitting function expression, and numerical computation was used to obtain the voltage gain at each stage of multistage CSVACs when determining the optimal number of stages.

When the relationship between the input and output voltage is simulated for the transistor voltage amplifier circuit, the supply voltage at the drain is set to $15V_T$, i.e., $V_{DD} = 15V_T$. The input voltage signal is set to $0.1\sin(\tau/2)V_T$, i.e., $V_{in} = 0.1\sin(\tau/2)V_T$, where $\tau$ is the simulation time.

When the relationship between input and output voltage is simulated for the CSVAC, the supply voltage at the drain is set to $15V_T$, i.e., $V_d = 15V_T$. The transistor electron escape rate is $0.2/\beta\hbar$, i.e., $\Gamma = 0.2/\beta\hbar$. The electron escape rate of the load resistor is $0.01/\beta\hbar$, i.e., $\Gamma_L = 0.01/\beta\hbar$. The input voltage signal is $-2.5\sin(\tau/3)V_T$, i.e., $V_{in} = -2.5\sin(\tau/3)V_T$. Without loss of generality, the initial value of output voltage is assumed to be governed by a uniform distribution in the MCMC simulation.

## 2.2 Voltage gain

In the CSVAC, the voltage gain is defined as the ratio of the output signal amplitude to the input signal amplitude, which is expressed as

$$G = A_{out}/A_{in}, \tag{1}$$

where $A_{out}$ is the output signal amplitude and $A_{in}$ is the input signal amplitude, with $A_{V_{out}} \geq A_{V_{in}} \geq 0$. Unless otherwise specified, the CSVAC voltage gain in this article refers to the CSVAC voltage gain at the state where the amplifier operates at the maximum voltage gain when the input signal amplitude and supply voltage are fixed.

## 2.3 Device models and experimental testing

In the experimental testing of multistage CSVACs, the model of PMOS transistors is BSS84, while the model of NMOS transistors is BSS138. Custom printed circuit boards were designed and assembled for modules of multistage CSVAC. Moreover, variable resistors were added between transistors and electrodes to adjust the voltage gain by analogously changing the transistor electron escape rate. Experimental testing was conducted under ambient conditions, utilizing an ARRAY 3400A signal generator to produce a sinusoidal input signal with a 5V amplitude and a 100Hz frequency. The load resistor is $100\,\mathrm{k\Omega}$. The multistage CSVAC was powered by an ITECH IT6322A DC power supply, while a Tektronix MDO4034C oscilloscope was employed to observe signal waveforms and measure data. The overall CSVAC voltage gain was set to 1.3 during experimental testing. The supply voltage at each stage of the CSVAC was fixed at 15V for experimental testing.

## 3 RESULTS

### 3.1 Transistor voltage amplifier model based on stochastic thermodynamic

As pivotal components of modern electronic devices, information processing circuits based on CMOS technology typically integrate N-channel metal-oxide-semiconductor (NMOS) and P-channel metal-oxide-semiconductor (PMOS) field-effect transistors to achieve functions such as the logical computation and signal amplification. The classical NMOS and PMOS structures, depicted in Figure 1A, are three-terminal devices comprising a source, drain, and gate, constructed from insulators, semiconductors, and metal electrodes. Due to the Coulomb blockade effect, transistors can be abstracted to single electron energy levels in the mesoscopic scale[33,34]. Electrodes can be abstracted to electron reservoirs with fixed chemical potentials, i.e., the driving force of electron transfer. Consequently, the physical processes during the operation of transistors can be abstracted to electron transfer processes within the circuit, whose dynamical model is shown in Figure 1B. The stochastic thermodynamic transistor model is detailed in supplemental materials and methods. In Figure 1B, $V_d$ is the drain voltage and $V_s$ is the source voltage. $\varepsilon_N$ and $\varepsilon_P$ are energy levels of NMOS transistors and PMOS transistors, while $\mu_d$ and $\mu_s$ are chemical potentials of the drain and source electron reservoirs, respectively. The relationship between transistor energy levels and voltages can be approximated as linear[35]. The transistor transfer characteristics can be adjusted by changing the linear relationship between transistor energy levels and voltages. For example, the transistor energy levels can be adjusted by changing the transistor gate input voltage. Supplemental materials and methods and Figure S1 provide a detailed discussion of the stochastic thermodynamic transistor current-voltage characteristics.

To validate the amplifier effect of the stochastic thermodynamic transistor model, a transistor voltage amplifier circuit adopting an NMOS transistor is illustrated in Figure 1C. In Figure 1C, $V_{DD}$ is the direct current (DC) supply voltage at the drain, $R_d$ is the drain resistor, $V_{in}$ is the input voltage, and $V_{out}$ is the output voltage. The abstract dynamical process projecting the operation of the transistor voltage amplifier circuit is shown in Figure 1D, where $\mu_{DD}$ is the chemical potential of the DC power source at the drain, and $\Gamma_r$ is the electron escape rate of the drain resistor. $\beta = 1/kT$, where $k$ is the Boltzmann constant and $T = 300\text{K}$ is the environment temperature. $\hbar$ is the approximate Planck constant. $V_T = kT/q \approx 26\text{mV}$ is the standard thermal voltage, i.e., the unit voltage of supply voltage, where $q = 1.6 \times 10^{-19}$ Coulomb is the electron charge. Based on Ohm's law and the transistor voltage amplifier model based on stochastic thermodynamics, a quantitative relationship between the resistance value and the electron escape rate of the drain resistor is derived, i.e., $R_d = 8.432/\Gamma_r$. There exists an inverse proportionality relationship between the resistance value and the electron escape rate of drain resistor. The detailed construction process of the transistor voltage amplifier model based on stochastic thermodynamics, as well as the relationship between the resistance value and the electron escape rate, is derived in supplemental materials and methods.

The relationships between the input and output voltages of transistor voltage amplifier circuits with different $\Gamma_r$ are shown in Figure 1E. Detailed simulation parameters are presented in the Materials and Methods. The relationships between the input and output voltages of the transistor voltage amplifier circuit can be effectively changed by adjusting $\Gamma_r$. The impact of the input voltage on the output voltage becomes more significant with the decrease of $\Gamma_r$. Figure 1F shows the output signal alternating component of the transistor voltage amplifier circuit when a sinusoidal signal is input. The simulation unit time is $\beta\hbar \approx 25\text{fs}$. For the same sinusoidal input signal, it can be observed that the output signal amplitudes increase with the decrease of $\Gamma_r$.

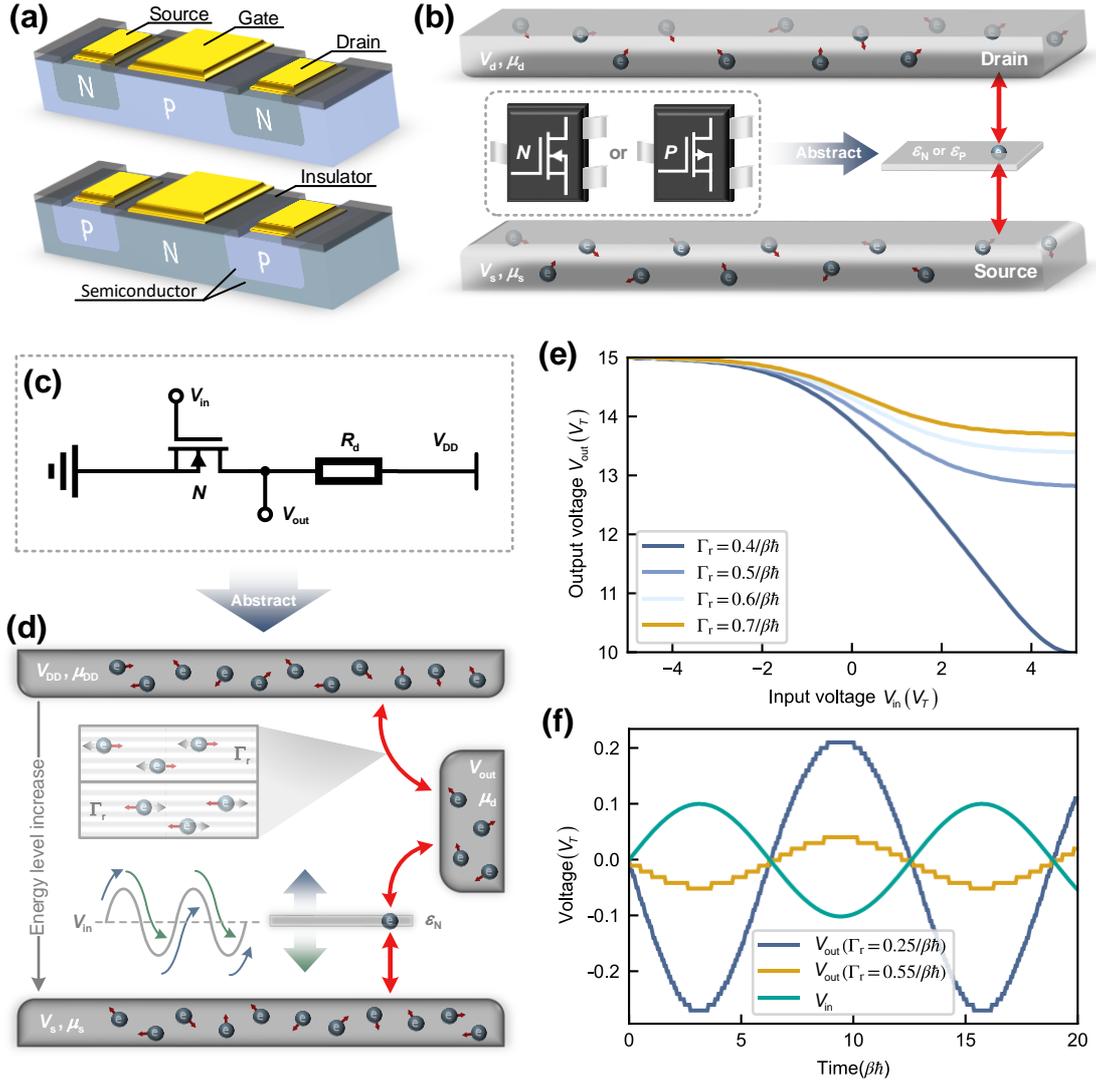

**Figure 1 Stochastic thermodynamic transistor model and amplifier effects. (a)** Classical NMOS (top) and PMOS (bottom) structures. **(b)** Dynamical model of the transistor operation process, where red unidirectional arrows denote the direction of electron motion, and red bidirectional arrows denote the electron transfer. **(c)** Transistor voltage amplifier circuit structure. **(d)** Dynamical model of transistor

voltage amplifier circuit operation process, where gray arrows denote impediments to electron transfer. As the input voltage increases, the energy level of NMOS transistors decreases. As the input voltage decreases, the energy level of NMOS transistors increases. **(e)** Relationships between the input and output voltages of transistor voltage amplifier circuit with different $\Gamma_r$. **(f)** Output signal alternating component when a sinusoidal signal is input into the transistor voltage amplifier circuit.

**3.2 CSVAC stochastic thermodynamic model**

The CSVAC is a typical amplifier circuit commonly applied for signal amplification. As shown in Figure 2A, the CSVAC structure includes an NMOS transistor and a PMOS transistor, where $R_L$ is the load resistor. The abstract dynamical process of the CSVAC operation process is shown in Figure 2B, where $\mu_{d_P}$, $\mu_{d_N}$, and $\mu_g$ are the chemical potentials of PMOS drain, NMOS drain and ground, respectively. Furthermore, the derivation of power dissipation of CSVACs is detailed in supplemental materials and methods.

Based on the CSVAC stochastic thermodynamic model, the relationships between input and output voltages of CSVAC are simulated by numerical and Markov Chain Monte Carlo (MCMC) simulations in Figure 2C. The simulation results indicate that the functionality of the CSVAC stochastic thermodynamic model is consistent with the functionality of traditional CSVAC. Detailed simulation parameters are presented in the Materials and Methods.

Based on the CSVAC stochastic thermodynamic model, the power dissipation with different input signal amplitudes and voltage gains (Equation (1), Materials and Methods) is shown in Figure 2D and Figure 2E, where $G$ is the voltage gain, and $A_{in}$ is the input signal amplitude. With the increase of voltage gain, the CSVAC power dissipation with the same input signal amplitude approximately exhibits exponential growth. Based on the analytical expression of CSVAC power dissipation in supplemental materials and methods, which quantifies the impact of voltage gain and input signal amplitude on the power dissipation, this phenomenon arises from two factors. Firstly, as the CSVAC voltage gain increases, a more significant transistor electron escape rate is required to maintain the same voltage gain level due to the nonlinearity between the transistor electron escape rate and the CSVAC voltage gain. It leads to an increase in CSVAC drain current and then ultimately results in higher power dissipation. Secondly, for the CSVAC shown in Figure 2A, to prevent the output signal distortion, more significant voltage gains necessitate higher supply voltages to stabilize the signal, which also leads to an increased power dissipation. To verify whether this phenomenon exists in physical circuits, a CSVAC physical entity circuit is made and experimentally tested. The CSVAC photograph, including physical entities and external connections, is shown in Figure 2F. Detailed information regarding the making and experimental test is presented in the Materials and Methods. The experimental results validate the nonlinear relationship among the power dissipation, the input signal amplitude and voltage gain in CSVACs, as shown in Figure 2 (G and H). Therefore, it is quantified for the inherent mechanism of power dissipation phenomenon, i.e., the CSVAC power dissipation exponentially increases with the increase of voltage gain. It is the most important theory basis to design a multistage amplifier scheme for reducing the power dissipation in CSVAC.

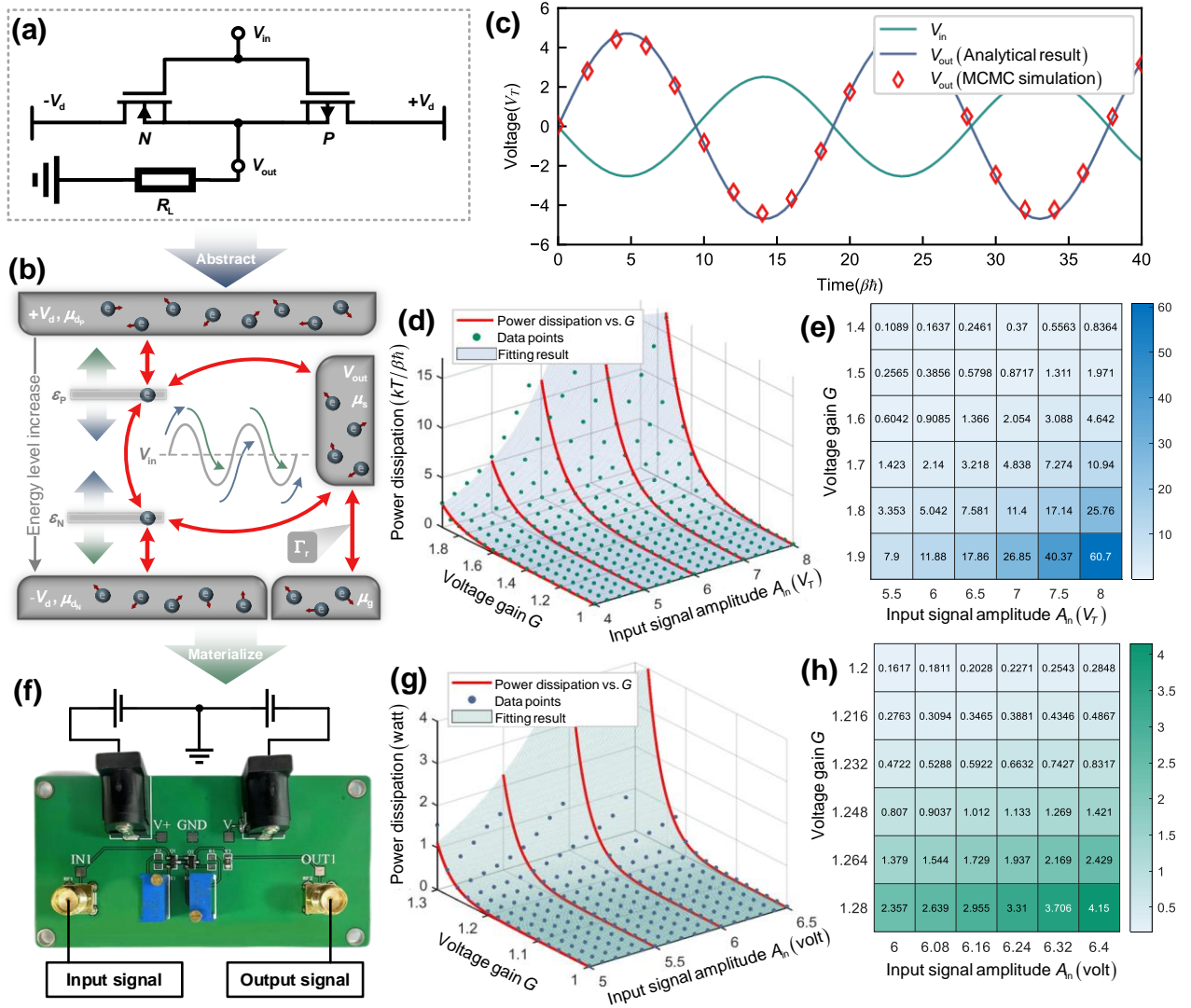

**Figure 2 Complementary Symmetry Voltage Amplifier Circuits (CSVAC). (a)** CSVAC structure. **(b)** Dynamical model of CSVAC operation process. **(c)** Relationships between input and output voltages of CSVAC, along with MCMC simulation results. **(d)** Relationships among the power dissipation, input signal amplitude, and voltage gain. The fit Root Mean Square Error (RMSE) is 0.06571 (closer to 0 indicates better fit), and the R-square is 0.9993 (ranging from 0 to 1, closer to 1 indicates stronger interpretability of fit results). **(e)** Heatmap of CSVAC power dissipation with different input signal amplitudes and voltage gains based on the stochastic thermodynamic model. **(f)** Physical entity photograph of CSVAC and external connections. **(g)** Relationships among the power dissipation, input signal

amplitude, and voltage gain, based on the CSVAC physical entity. The fit RMSE is 0.008732, and the R-square is 0.9983. (**h**) Heatmap of CSVAC power dissipation with different input signal amplitudes and voltage gains based on the physical entity circuit.

**3.3 Multistage CSVAC architecture**

The multistage CSVAC architecture is shown in Figure 3A. In Figure 3A, the functionality of the single stage CSVAC with the voltage gain $G$ can be realized by the multistage CSVAC with stage voltage gains $G_\lambda$, where $\sum_{\lambda=1}^{n} G_\lambda = G$, $G \geq G_\lambda \geq 1$. In Figure 3A, for $\lambda = 1, 2, ..., n$, $V_{d_\lambda}$ is the drain voltage of the $\lambda$ th stage in the multistage CSVAC, $P_\lambda$ and $N_\lambda$ denote PMOS transistors and NMOS transistors of the $\lambda$ th stage in the multistage CSVAC, respectively. In the multistage CSVAC architecture, the power dissipation of multistage CSVAC can be summarized by the power dissipation of each stage CSVAC. An optimal multistage scheme is proposed in this article to reduce the multistage CSVAC power dissipation while keeping the overall voltage gain constant.

When the number of stages is configured as $K$ in the multistage CSVAC, it is proven that the total multistage CSVAC power dissipation is a convex function of voltage gains at each stage except for the last one, i.e., $G_1, G_2, ..., G_{K-1}$. For example, the total two-stage CSVAC power dissipation with an overall voltage gain of 1.5 is a convex function of the first stage voltage gain, i.e., $G_1$, as shown in Figure 3B. To eliminate discrepancies in the total two-stage CSVAC power dissipation values caused by different input signal amplitudes, the total two-stage CSVAC power dissipation has been normalized in Figure 3B. Based on results in Figure 3B, the optimal voltage gain $G_1^{\text{opt}}$ that minimizes the total two-stage CSVAC power dissipation lies between the maximum and minimum voltage gains, i.e., $G > G_1^{opt} > 1$. Results in Figure 3B indicate that the optimal voltage gain $G_1^{opt}$ gradually approaches one as the input signal amplitude increases. This result implies that the input signal should not be amplified at the first stage of two-stage CSVAC. For a multistage CSVAC with $K$ stages and an overall voltage gain $G$, it is proven that the total power dissipation of multistage CSVAC is minimized when $G_\lambda$ satisfies $G_\lambda = G^{1/K}$. When the overall voltage gain of multistage CSVAC exceeds $1.075^K$, i.e., $G > 1.075^K$, the multistage CSVAC with $K$ stages has a lower total power dissipation than the total power dissipation of multistage CSVAC with the number of stages less than $K$. In this case, the multistage CSVAC with the number of stages as $K$ is considered better than those with fewer stages. The detailed proof of relevant conclusions of multistage CSVACs is provided in supplemental materials and methods. Since the multistage CSVAC architecture requires more components, it leads to additional power dissipation. When the multistage CSVAC with $K$ stages satisfies $P_{sum}^{(\kappa)}(\cdot) < P_{sum}^{(\kappa-1)}(\cdot)$, where $\kappa \in \{2, 3, ..., K\}$, the number of stages is considered the optimal number of stages, denoted as $K_{opt}$. Therefore, the optimal multistage scheme can be illustrated as shown in Scheme 1.

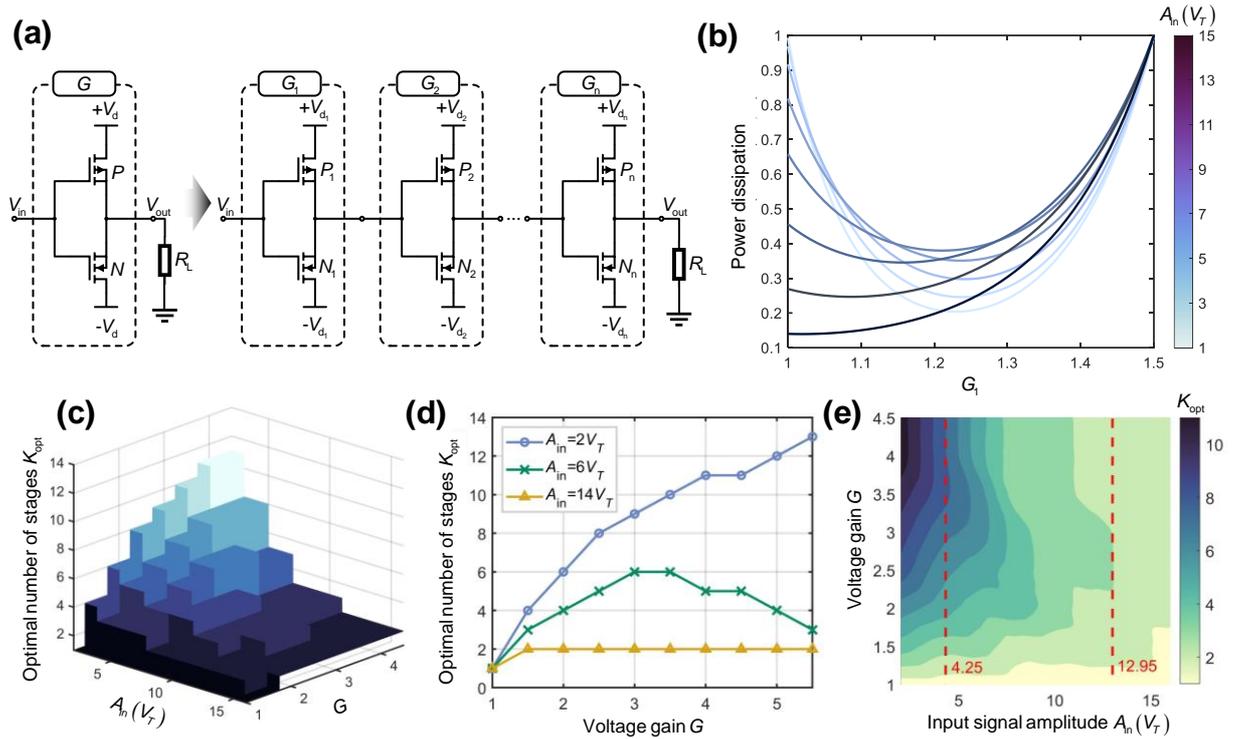

**Figure 3 Multistage CSVAC.** (**a**) Single stage CSVAC vs multistage CSVAC. (**b**) Total power dissipation of the two-stage CSVAC with respect to the first stage voltage gain. (**c**) Optimal number of stages for the multistage CSVAC with different input signal amplitudes and voltage gains. (**d**) Relationship between the optimal number of stages and voltage gain with different input signal amplitudes. (**e**) Contour plot of the optimal number of stages for the multistage CSVAC with different input signal amplitudes and voltage gains.

For a multistage CSVAC, the optimal number of stages with respect to different input signal amplitudes and voltage gains is shown in Figure 3C. When the voltage gain is fixed in multistage CSVACs, the optimal number of stages decreases with the increase of input signal amplitude. However, the relationship between the optimal number of stages and voltage gain is complex considering different input signal amplitudes. To illustrate this point more clearly, Figure 3C provides examples with $A_{in} = 2V_T$, $A_{in} = 6V_T$, and $A_{in} = 14V_T$, showing the relationship between the optimal number of stages and voltage gain. When the input signal amplitude is small in multistage CSVACs, e.g., $A_{in} = 2V_T$, the optimal number of stages increases with the increase of voltage gain. Based on the power dissipation of multistage CSVACs in supplemental materials and methods, the increasing voltage gain is the main factor for the increasing power dissipation when the input signal amplitude is small, e.g. $A_{in} = 2V_T$. When the input signal amplitude is large in multistage CSVACs, e.g., $A_{in} = 14V_T$, the optimal number of stages is independent of the voltage gain when the voltage gain exceeds 1.5. Considering the power dissipation of multistage CSVACs in supplemental materials and methods, the reason is that the impact of the voltage gain on the power dissipation is gradually diminished in the multistage CSVAC when the input signal amplitude exceeds 1.5. Between these two cases, e.g., $A_{in} = 6V_T$, a "transition phase" exists, where the optimal number of stages initially increases with the increase of voltage gain and then decreases with the increase of voltage gain when the voltage gain exceeds 3.5. During the "transition phase", the optimal number of stages initially increases with the increase of voltage gain because the multistage architecture still optimizes the CSVAC power dissipation. It is important to note that the subsequent decrease of the optimal number of stages with increasing voltage gain does not necessarily mean that decreasing the number of stages reduces the power dissipation of multistage CSVACs. Due to limited data precision, the change of power dissipation is less than the precision in Figure 3D when the number of stages increases. Therefore, this change part is ignored, and a smaller number of stages is selected as the optimal number of stages in Figure 3D. To obtain a more accurate depiction of "transition phase", we increase the simulation precision to obtain a contour plot of the optimal number of stages for multistage CSVACs with different input signal amplitudes and voltage gains, as shown in Figure 3E. Based on results in Figure 3E, it can be observed that the optimal number of stages exists a "transition phase" when the input signal amplitude is between 4.25 $V_T$ and 12.95 $V_T$ in the multistage CSVAC, where the optimal number of stages initially increases and then decreases with the increase of voltage gain.

| Scheme 1: Optimal multistage scheme |
| --- |
| Input: Multistage CSVAC power dissipation $P_{\text{sum}}^{(K)}(\cdot)$ and overall voltage gain $G$ |
| Output: Optimal number of stages $K_{\text{opt}}$ and optimal voltage gains at each stage $\mathbf{G}_{K_{\text{opt}}}^{\text{opt}} = \left[ G_1^{\text{opt}}, G_2^{\text{opt}}, ..., G_{K_{\text{opt}}}^{\text{opt}} \right]$ |
| 1:    Calculate $\mathbf{G}_2^{\text{opt}}$; |
| 2:    if $P_{\text{sum}}^{(1)}(G) \leq P_{\text{sum}}^{(2)}\left(\mathbf{G}_2^{\text{opt}}\right)$ |
| 3:    $K_{\text{opt}} = 1$, $\mathbf{G}_{K_{\text{opt}}}^{\text{opt}} = \mathbf{G}_1^{\text{opt}} = \left[G_1^{\text{opt}}\right] = G$; |
| 4:    else |
| 5:    $K = 2$; |
| 6:    while $P_{\text{sum}}^{(K-1)}\left(\mathbf{G}_{K-1}^{\text{opt}}\right) > P_{\text{sum}}^{(K)}\left(\mathbf{G}_K^{\text{opt}}\right)$ |
| 7:    $K_{\text{opt}} = K$, $\mathbf{G}_{K_{\text{opt}}}^{\text{opt}} = \mathbf{G}_K^{\text{opt}}$; |
| 8:    $K = K + 1$; |
| 9:    Calculate $\mathbf{G}_K^{\text{opt}}$; |
| 10:  endwhile |
| 11:  endif |

Based on the CSVAC stochastic thermodynamic model, the power dissipation of single stage amplification and the power dissipation of optimal multistage amplification are compared for CSVACs with different input signal amplitudes and voltage gains in Figure 4A. When the input signal amplitude is fixed in CSVACs, simulation results in Figure 4B show the saved power dissipation between the single stage CSVAC and optimal multistage CSVAC increases with the increase of voltage gain. When the voltage gain is fixed in CSVACs, simulation results in Figure 4C show that the saved power dissipation between the single stage CSVAC and optimal multistage CSVAC decreases with the increase of input signal amplitude. Across various input signal amplitudes and voltage gains, multistage architecture consistently optimizes the power dissipation of the multistage CSVAC. Furthermore, the effectiveness of optimization becomes more significant as the input signal amplitude decreases and the voltage gain increases. In particular, at the boundaries of the variable range, i.e., the input signal amplitude is $2V_T$ and the voltage gain is 2, the optimal multistage architecture results in a 99.36% reduction in the power dissipation for multistage CSVAC compared with the power dissipation of traditional CSVAC. When the voltage gain is configured with a specified threshold (detailed in Materials and Methods), a physical entity of multistage CSVACs is made and experimentally tested to validate the optimization effect of multistage CSVAC architecture on power dissipation, as shown in Figure 4D. Details of device models and experiment configurations are provided in the Materials and Methods. The optimal number of stages and the optimal voltage gain at each stage are detailed in supplemental materials and methods for the physical entity of multistage CSVACs. And the relationship between the minimum power dissipation and the number of stages of physical multistage CSVACs is shown in Figure S2. Based on experimental testing results, the power dissipation values of the single stage CSVAC and multistage CSVAC with different voltage gains are shown in Figure 4 (E, F and G). Compared with the power dissipation of traditional single stage CSVAC, the power dissipation of two-stage CSVAC is reduced by 86.49%, 94.59%, and 91.89% when the first stage voltage gain is 1.06, 1.16, and 1.21, respectively. It is noteworthy that the optimal stage voltage gain is employed for the two-stage CSVAC to achieve the maximum power dissipation reduction in Figure 4F.

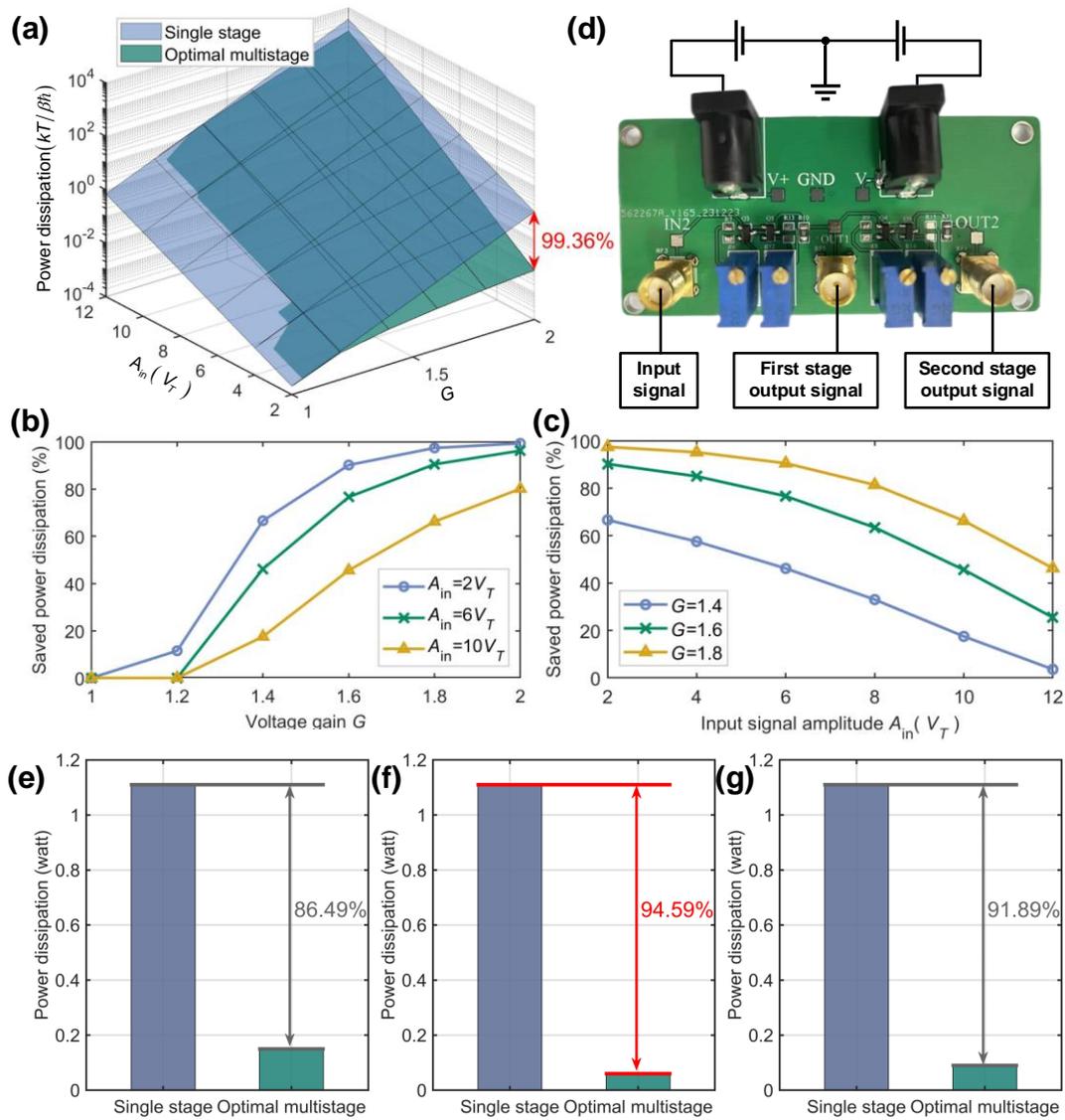

**Figure 4 Power dissipation optimization of the multistage CSVAC. (a)** Power dissipation of the single stage amplification and the optimal multistage amplification

for the CSVAC with different input signal amplitudes and voltage gains, based on the stochastic thermodynamic model. **(b)** Saved power dissipation between the single stage CSVAC and optimal multistage CSVAC when the input signal amplitude is fixed. **(c)** Saved power dissipation between the single stage CSVAC and optimal multistage CSVAC when the voltage gain is fixed. **(d)** Physical entity photograph of the two-stage CSVAC and external connections. **(e)**, **(f)**, and **(g)**, Power dissipation of the single stage amplification and the optimal multistage amplification for the two-stage CSVAC with different voltage gains ( $G_1 = 1.06$ , $G_1 = G_1^{opt} = 1.16$ , and $G_1 = 1.21$ , respectively), based on the physical entity circuit.

## 4 CONCLUSION

Based on stochastic thermodynamics, a stochastic model of single-level transistors, which is able to describe the electron transfer process consistent with the fluctuation theorem while guaranteeing compatibility with the CMOS infrastructure, is proposed in this article. Taking CSVAC as an example, we utilize the single-level transistor model as a new insight to quantify the nonlinear relationship among the power dissipation, the input signal amplitude and voltage gain in the CSVAC. Based on the power dissipation analysis of the CSVAC stochastic thermodynamic model, the inherent mechanism of exponential growth power dissipation is first clearly explained in this article. Furthermore, a multistage CSVAC architecture is proposed to reduce the CSVAC power dissipation. Compared with the power dissipation of traditional CSVAC, simulation results show that the optimal multistage scheme proposed in this article can reduce the CSVAC power dissipation by up to 99.36%. Moreover, a physical entity of multistage CSVACs is made and experimentally tested to validate the optimization effect of multistage CSVAC architecture on power dissipation and reduced the CSVAC power dissipation by up to 94.59% compared with the power dissipation of traditional CSVAC. Future work will explore the integration of the multistage architecture into more complex amplifier circuits and investigate its potential for enhancing the overall efficiency of electronic systems.

## RESOURCE AVAILABILITY

### Data and code availability

Data are available from the corresponding author upon reasonable request.

## FUNDING AND ACKNOWLEDGMENTS

This work was supported by the National Nature Science Foundation of China grant 62120106007.

## AUTHOR CONTRIBUTIONS

Xiaoxuan Peng: Methodology, Investigation, Visualization, Writing―original draft, Writing―review & editing.

Xiaohu Ge: Conceptualization, Methodology, Investigation, Supervision, Writing―review & editing.

## DECLARATION OF INTERESTS

Xiaoxuan Peng and Xiaohu Ge declare that they have no conflict of interest.

## SUPPLYMENTAL INFORMATION

Document S1. Figures S1–S2, and supplemental materials and methods.

**LEAD CONTACT WEBSITE**

http://eic.hust.edu.cn/professor/gexiaohu/index.html

30. Ge, X., Ruan, M., Peng, X. et al. (2024). An energy and information analysis method of logic gates based on stochastic thermodynamics. *PNAS Nexus*. 3:pgae365.

31. Sharroush, S.M. (2019). Design of the CMOS inverter-based amplifier: A quantitative approach. *Int. J. Circuit Theory Appl*. 47:1006-1036.

32. Liu, L., Zou, J., Ma, N. et al. (2015). A CMOS transimpedance amplifier with high gain and wide dynamic range for optical fiber sensing system. *Optik*. 126:1389-1393.

33. Sahu, S.K. and Mazumdar, K. (2024). State-of-the-art analysis of single electron transistor. In *2024 IEEE 3rd International Conference on Electrical Power and Energy Systems (ICEPES)*, pp. 1-6. IEEE.

34. Hussain, N., Hossain, M.A., Mumenin, N. et al. (2022). Characteristics of energy in the Coulomb island of single electron transistor. In *2022 International Conference on Recent Progresses in Science, Engineering and Technology (ICRPSET)*, pp. 1-5. IEEE.

35. Datta, S. (1997). Electronic Transport in Mesoscopic Systems. Cambridge University Press, Cambridge.

# Supplemental information

## Modeling and Optimization of Transistor Voltage Amplifiers Based on Stochastic Thermodynamics

Xiaoxuan Peng, Xiaohu Ge

# SUPPLEMENTAL INFORMATION

# MATERIALS AND METHODS

## Stochastic Thermodynamic Transistor Model

Let $q$ denote an electron charge, $V_{in}$ is the input voltage and the energy level of NMOS transistors is expressed as

$$\varepsilon_N = \varepsilon_N^0 - qV_{in}, \tag{S1}$$

where $\varepsilon_N^0$ is the reference energy of NMOS transistors. The energy level of PMOS transistors is expressed as

$$\varepsilon_P = \varepsilon_P^0 + qV_{in}, \tag{S2}$$

where $\varepsilon_P^0$ is the reference energy of PMOS transistors. The chemical potential of drain electron reservoir is expressed as

$$\mu_d = -qV_d, \tag{S3}$$

where $V_d$ is the drain voltage. The chemical potential of source electron reservoir is configured as

$$\mu_s = 0. \tag{S4}$$

$k_{ij}$ and $k_{ji}$ are the transfer rates for electrons from the transistor $j$ to the electrode $i$ and electrons from the electrode $i$ to the transistor $j$, respectively, $j \in \{P, N\}$ and $i \in \{d, s\}$. To guarantee the thermodynamic consistency, the transfer rates satisfy the local detailed balance, which is the prerequisite of stochastic thermodynamics[1], i.e.,

$$\frac{k_{ji}}{k_{ij}} = e^{-\beta(E_j - E_i)}, \tag{S5}$$

where $E_j$ is the energy described by the energy level of transistor $j$, $E_i$ is the chemical potential of electrode $i$. $\beta = 1/kT$, where $k$ is the Boltzmann constant and $T$ is the environment temperature in Kelvin. The local detailed balance ensures the described electron transfer process is consistent with the fluctuation theorems. Based on the distribution characteristics of identical particles, the electron transfer rate between electrodes and transistors is expressed as[2]

$$\begin{cases} k_{ji} = \Gamma f_i(\varepsilon_j) \\ k_{ij} = \Gamma \left[1 - f_i(\varepsilon_j)\right] \end{cases}, \tag{S6}$$

where $f_i(x) = \left[e^{\beta(x-\mu_i)} + 1\right]^{-1}$ is the Fermi-Dirac distribution, $\Gamma$ is the transistor electron escape rate.

Based on the Coulomb blocking effect, a single energy level can accommodate at most one electron, with the electron number $n_j$ of the transistor $j$ being either 0 or 1. Taking NMOS transistors as an example, the Markov equation of NMOS transistor states is expressed as

$$dp_{n_N}^t / dt = \mathbf{R}_{\text{NMOS}} \times p_{n_N}^t, \tag{S7}$$

where $\mathbf{R}_{\text{NMOS}}$ is the transition rate matrix of NMOS transistors and $p_{n_N}^t$ is the probability distribution of NMOS transistors at time $t$. The transition rate matrix of NMOS transistors is expressed as

$$\mathbf{R}_{\text{NMOS}} = \begin{bmatrix} -F_1 & k_{\text{dN}} + k_{\text{sN}} \\ k_{\text{Nd}} + k_{\text{Ns}} & -F_2 \end{bmatrix}, \tag{S8}$$

where $F_n = \sum_{m \neq n} (\mathbf{R}_{\text{NMOS}})_{m,n}$ with subscripts $m$ and $n$ denoting the matrix row and column numbers, respectively. Based on the steady state distribution of transition rate matrix, the average number of electrons $\langle n_N \rangle$ in the NMOS transistor is expressed as

$$\langle n_N \rangle = \frac{\text{Null}(\mathbf{R}_{\text{NMOS}})_{1,0}}{\sum_{m=0}^{1} \text{Null}(\mathbf{R}_{\text{NMOS}})_{m,0}}, \tag{S9}$$

where $\text{Null}(\cdot)$ is a null space matrix with subscripts denoting the element row and column numbers. The current $J_{d \to N}$ from the drain to the NMOS transistor is expressed as

$$J_{d \to N} = q \left[ k_{Nd} (1 - \langle n_N \rangle) - k_{dN} \langle n_N \rangle \right]. \tag{S10}$$

**Stochastic Thermodynamic Transistor Current-Voltage Characteristics**

Without loss of generality, in this article, the simulation unit time is set to $\beta\hbar \approx 25\text{fs}$, where $\hbar$ is the approximate Planck constant. The environment temperature is $T = 300\text{K}$. The thermal voltage is $V_T = kT/q \approx 26\text{mV}$, which is the unit voltage of supply voltage. As a consequence, the supply voltage is denoted as $V_d = 15V_T$. The transistor electron escape rate is $\Gamma = 0.2/\beta\hbar$. Based on the stochastic thermodynamic transistor model, the drain current $i_D^{\text{NMOS}}$ of NMOS transistors with different gate voltages $V_{\text{in}}$, i.e., the transfer characteristics of NMOS transistors, is shown in Figure S1A. Based on the transfer characteristics, it can be observed that NMOS transistors have a pinch-off voltage $V_{PN}$, where $V_{PN}$ increases with the increase of reference energy $\varepsilon_N^0$. By setting the reference energy of NMOS transistors, its characteristics can be made to conform to characteristics of enhancement-type NMOS transistors, depletion-type NMOS transistors, and depletion-type N-channel junction field-effect transistors (JFET), etc. In this article, the depletion-type NMOS transistors is used as an example without special instructions. When $\varepsilon_N^0 = 0$, the pinch-off voltage is $V_{PN} = -5V_T$. When $V_{in} < V_{PN}$, the NMOS transistor is cut off and the drain current is approximately 0.

When $\varepsilon_N^0 = 0$, the drain current $i_D^{NMOS}$ of NMOS transistors with different drain-source voltages $V_{DS}^{NMOS}$, i.e., the output characteristic of NMOS transistors, is shown in Figure S1B. The red dashed line in Figure S1B is the pre-pinch-off trajectory, i.e., $V_{DS}^{NMOS} = V_{in} - V_{PN}$. The left side of red dashed line corresponds to the non-saturation region of NMOS transistors, where $i_D^{NMOS}$ increases with the increase of $V_{DS}^{NMOS}$. When $V_{in}$ increases, the increase of $i_D^{NMOS}$ becomes more significant, which indicates the variable resistor characteristic of NMOS transistors. On the right side of red dashed line is the saturation region of NMOS transistors, where $V_{DS}^{NMOS}$ increases but $i_D^{NMOS}$ remains constant. When $V_{in}$ increases, $i_D^{NMOS}$ also increases, which indicates the amplifier effect of NMOS transistors.

Based on the stochastic thermodynamic transistor model, the drain current $i_D^{PMOS}$ of PMOS transistors with different gate voltages $V_{in}$, i.e., the transfer characteristics of PMOS transistors, is shown in Figure S1C. Based on transfer characteristics, it can be observed that PMOS transistors have a pinch-off voltage $V_{PP}$, where $V_{PP}$ increases with the decrease of reference energy $\varepsilon_P^0$. By setting the reference energy of PMOS transistors, its characteristics can be made to conform to characteristics of enhancement-type PMOS transistors, depletion-type PMOS transistors, and depletion-type P-channel JFETs, etc. In this article, the depletion-type PMOS transistor is used as an example without special instructions. When $\varepsilon_P^0 = 0$, the pinch-off voltage is $V_{PP} = 5V_T$. When $V_{in} > V_{PP}$, the PMOS transistor is cut off and the drain current is approximately 0.

When $\varepsilon_P^0 = 0$, the drain current $i_D^{PMOS}$ of NMOS transistors with different drain-source voltages $V_{DS}^{PMOS}$, i.e., the output characteristic of PMOS transistors, is shown in Figure S1D. The red dashed line in Figure S1D is the pre-pinch-off trajectory, i.e., $V_{DS}^{PMOS} = V_{PP} - V_{in}$. The left side of red dashed line corresponds to the non-saturation region of PMOS transistors, where $i_D^{PMOS}$ increases with the increase of $V_{DS}^{PMOS}$. When $V_{in}$ decreases, the increase of $i_D^{PMOS}$ becomes more significant, which indicates the variable resistor characteristic of PMOS transistors. On the right side of red dashed line is the saturation region of PMOS transistors, where $V_{DS}^{PMOS}$ increases but $i_D^{PMOS}$ remains constant. When $V_{in}$ increases, $i_D^{PMOS}$ also increases, which indicates the amplifier effect of PMOS transistors.

**Transistor Voltage Amplifier Model Based on Stochastic Thermodynamics**

Let $V_{DD}$ be the direct current (DC) supply voltage at the drain, $R_d$ is the drain resistor, and $V_{out}$ is the output voltage. The energy level of NMOS transistors is expressed as

$$\varepsilon_N = -qV_{in}. \tag{S11}$$

The chemical potential of drain electron reservoir of NMOS transistors is expressed as

$$\mu_d = -qV_{out}. \tag{S12}$$

The chemical potential of DC power supply at the drain is expressed as

$$\mu_{DD} = -qV_{DD}. \tag{S13}$$

The chemical potential of source electron reservoir is configured as

$$\mu_s = 0. \tag{S14}$$

During each time interval $\Delta t$ of transistor voltage amplifier circuit operations, the output voltage $V_{out}$ is changed, which results in the change of electron transfer rates and the update of transition rate matrix. Based on the steady state distribution of transition rate matrix, the average number of electrons can be derived for the transistor.

The current from the DC power supply to the drain resistor is expressed as

$$J_{DD \to R_d} = q \frac{\Gamma_r}{2} \left[ f_d(\mu_d) - f_{DD}(\mu_d) \right], \tag{S15}$$

where $\Gamma_r$ is the electron escape rate of resistors and $f_{DD}(\mu_d) = \left[ e^{\beta(\mu_d - \mu_{DD})} + 1 \right]^{-1}$. The current from the electrode $i$ to the transistor $j$ is expressed as

$$J_{i \to j} = q \left[ k_{ji} \left( 1 - \langle n_j \rangle \right) - k_{ij} \langle n_j \rangle \right], \tag{S16}$$

where $\langle n_j \rangle$ is the average number of electrons in the transistor $j$. When

$$J_{DD \to R_d} = J_{d \to N}, \tag{S17}$$

the transistor voltage amplifier circuit reaches a steady state and the output voltage $V_{out}$ can be determined.

The role of resistors is to impede the electron transfer. Based on Ohm's law, the output voltage $V_{out}$ at the drain can be expressed as

$$V_{out} = V_{DD} - J_{d \to N} \cdot R_d. \tag{S18}$$

Based on (S18) and the stochastic thermodynamic model of transistor voltage amplifier circuit, the relationship between the resistance value and the electron escape rate is approximately expressed as

$$R_d = 8.432/\Gamma_r. \tag{S19}$$

The relationship between the electron escape rate and resistance approximately exhibits an inversely proportional relationship, which aligns with the conventional understanding.

**Power Dissipation of complementary symmetric voltage amplifier circuits (CSVACs)**

Let $R_L$ be the load resistance. The energy level of PMOS transistors is expressed as

$$\varepsilon_P = qV_{in}, \tag{S20}$$

The energy level of NMOS transistors is expressed as

$$\varepsilon_N = \varepsilon_N^0 - qV_{in}, \tag{S21}$$

where $\varepsilon_N^0 = qV_d$. The chemical potential of drain electron reservoir of PMOS transistors is expressed as

$$\mu_{d_P} = -qV_d. \tag{S22}$$

The chemical potential of drain electron reservoir of NMOS transistors is expressed as

$$\mu_{d_N} = qV_d. \tag{S23}$$

The chemical potential of source electron reservoir is expressed as

$$\mu_s = -qV_{out}. \tag{S24}$$

The chemical potential of power ground is configured as

$$\mu_g = 0. \tag{S25}$$

The electron transfer rate between PMOS transistors and NMOS transistors depends on the relative energy levels. $k_{PN}$ is the transfer rates for electrons from NMOS transistors to PMOS transistors. $k_{NP}$ is the transfer rates for electrons from PMOS transistors to NMOS transistors. Based on the distribution characteristics of electron pairs, $k_{PN}$ and $k_{NP}$ are respectively expressed as[2]

$$k_{PN} = \begin{cases} \Gamma o(\varepsilon_P - \varepsilon_N), & \varepsilon_P > \varepsilon_N \\ \Gamma[1 + o(\varepsilon_N - \varepsilon_P)], & \varepsilon_P \leq \varepsilon_N \end{cases}, \quad (S26)$$

and

$$k_{NP} = \begin{cases} \Gamma[1 + o(\varepsilon_P - \varepsilon_N)], & \varepsilon_P > \varepsilon_N \\ \Gamma o(\varepsilon_N - \varepsilon_P), & \varepsilon_P \leq \varepsilon_N \end{cases}, \quad (S27)$$

where $o(x) = [e^{\beta x} - 1]^{-1}$ is the Bose-Einstein distribution.

The current flowing through the load resistor is expressed as

$$J_{CSVAC \to R_L} = \frac{\Gamma_L}{2}[f_g(\mu_s) - f_s(\mu_s)], \quad (S28)$$

where $\Gamma_L$ is the electron escape rate of load resistor and $f_g(\mu_s) = [e^{\beta(\mu_s - \mu_g)} + 1]^{-1}$. When

$$J_{CSVAC \to R_L} = J_{P \to s} + J_{N \to s}, \quad (S29)$$

the entire circuit, including the CSVAC and the load, reaches a steady state and the output voltage of CSVACs can be determined.

The CSVAC power dissipation arises from the conjugate affinity between transistors and electrodes, which can be calculated by the product of current and potential differences[3]. Therefore, the CSVAC power dissipation is analytically expressed as

$$P_{CSVAC} = P_{PMOS} + P_{NMOS}, \tag{S30A}$$

$$P_{PMOS} = J_{d_P \to P}(\mu_s - \mu_{d_P}), \tag{S30B}$$

$$P_{NMOS} = J_{d_N \to N}(\mu_s - \mu_{d_N}), \tag{S30C}$$

$$J_{d_P \to P} = q\left[k_{Pd_P}\left(1 - \langle n_P \rangle\right) - k_{d_P P}\langle n_P \rangle\right], \tag{S30D}$$

$$J_{d_N \to N} = q\left[k_{Nd_N}\left(1 - \langle n_N \rangle\right) - k_{d_N N}\langle n_N \rangle\right], \tag{S30E}$$

where $k_{Pd_P} = \Gamma f_{d_P}(\varepsilon_P)$ is the transfer rate for electronics from drain electron reservoir of PMOS transistors to PMOS transistors, $k_{d_P P} = \Gamma\left[1 - f_{d_P}(\varepsilon_P)\right]$ is the transfer rate for electronics from PMOS transistors to drain electron reservoir of PMOS transistors, $f_{d_P}(\varepsilon_P) = \left[e^{\beta(\varepsilon_P - \mu_{d_P})} + 1\right]^{-1}$. $k_{Nd_N} = \Gamma f_{d_N}(\varepsilon_N)$ is the transfer rate for electronics from drain electron reservoir of NMOS transistors to NMOS transistors and $k_{d_N N} = \Gamma\left[1 - f_{d_N}(\varepsilon_N)\right]$ is the transfer rate for electronics from NMOS transistors to drain electron reservoir of NMOS transistors, $f_{d_N}(\varepsilon_N) = \left[e^{\beta(\varepsilon_N - \mu_{d_N})} + 1\right]^{-1}$.

**Proof of Multistage CSVACs**

Let $A_{in}$ be the input signal amplitude and $G$ be the total voltage gain of multistage CSVACs, respectively. Based on the regression analysis results of simulation data, the CSVAC power dissipation could be expressed as a function of $A_{in}$ and $G$, i.e.,

$$P_{\text{CSVAC}}(A_{\text{in}}, G) = \exp(a + bA_{\text{in}} + cG), \tag{S31A}$$

where

$$\begin{cases} a = -18.7 \\ b = 0.8156 \\ c = 8.569 \end{cases}. \tag{S31B}$$

For example, considering a multistage CSVAC with the number of stages as 2, the power dissipation of multistage CSVACs is expressed as

$$P_{\text{sum}}^{(2)} = P_{\text{CSVAC}}(A_{\text{in}}, G_1) + P_{\text{CSVAC}}(A_{\text{in}} G_1, G/G_1), \tag{S32}$$

where $G_1$ is the voltage gain of the first-stage CSVAC. When $A_{\text{in}}$ and $G$ are fixed, $P_{\text{sum}}^{(2)}$ is a function of $G_1$, i.e., $P_{\text{sum}}^{(2)}(G_1)$. $P_{\text{CSVAC}}(A_{\text{in}}, G)$ is a convex function of $G$ when $A_{\text{in}}$ is fixed. When $A_{\text{in}}$ and $G$ are fixed, $P_{\text{sum}}^{(2)}(G_1)$ is also a convex function of $G_1{}^4$. This conclusion can also be proved based on properties of convex functions, as follows:

When $0 \le \theta \le 1$, there is

$$G_1 = \theta + (1-\theta)G. \tag{S33}$$

To prove that $P_{\text{sum}}^{(2)}(G_1)$ is a convex function of $G_1$, it suffices to show that for $\forall \theta \in [0,1]$, there is

$$P_{\text{sum}}^{(2)}(\theta + (1-\theta)G) \le \theta P_{\text{sum}}^{(2)}(1) + (1-\theta) P_{\text{sum}}^{(2)}(G). \tag{S34}$$

Substituting (S31A) and (S32) into (S34), there is

$$\exp(a + bA_{\text{in}} + c(\theta + (1-\theta)G)) + \exp(a + bA_{\text{in}}(\theta + (1-\theta)G) + \frac{cG}{\theta + (1-\theta)G})$$
$$\le \begin{cases} \theta \exp(a + bA_{\text{in}} + c) + \theta \exp(a + bA_{\text{in}} + cG) \\ + (1-\theta)\exp(a + bA_{\text{in}} + cG) + (1-\theta)\exp(a + bA_{\text{in}} G + c) \end{cases}. \tag{S35}$$

Simplifying (S35) and letting $u = G - 1$, there is

$$\left\{ \begin{array}{l} \theta + \exp(cu) + (1-\theta)\exp(ubA_{in}) \\ -\exp(cu(1-\theta)) - \exp((1-\theta)ubA_{in} + \dfrac{cu\theta}{(1-\theta)u+1}) \end{array} \right\} \geq 0. \tag{S36}$$

Let

$$f(u, A_{in}) = \left\{ \begin{array}{l} \theta + \exp(cu) + (1-\theta)\exp(ubA_{in}) \\ -\exp(cu(1-\theta)) - \exp((1-\theta)ubA_{in} + \dfrac{cu\theta}{(1-\theta)u+1}) \end{array} \right\}, \tag{S37}$$

there is

$$\left\{ \begin{array}{l} f(0,0) = 0 \\ \dfrac{\partial f(u, A_{in})}{\partial u} \geq 0. \\ \dfrac{\partial f(u, A_{in})}{\partial A_{in}} \geq 0 \end{array} \right. \tag{S38}$$

Therefore, (S34) is established. For a multistage CSVAC with the number of stages as 2, the power dissipation is a convex function of the first-stage voltage gain in CSVACs if (S34) is satisfied.

To achieve the minimum power dissipation of a two-stage CSVAC with fixed $A_{in}$ and $G$, the necessary and sufficient condition is

$$\begin{aligned} \dfrac{\partial P_{sum}^{(2)}(G_1)}{\partial G_1} &= c\exp(a + bA_{in} + cG_1) + \exp(a + bA_{in}G_1 + \dfrac{cG}{G_1})(bA_{in} - \dfrac{cG}{G_1^2}) \\ &= 0 \end{aligned} \tag{S39}$$

Let $v = \dfrac{bA_{in}}{c}$, (S39) can be simplified to

$$\exp((v-1)G_1 - v + \dfrac{G}{G_1}) = \dfrac{G_1}{G/G_1 - vG_1}. \tag{S40}$$

Assuming $A_{in}$ is small enough such that $v \doteq 0$ when $c \gg b$, equation (S40) is equivalent to

$$G = G_1^2. \tag{S41}$$

To ensure that the power dissipation of two-stage CSVAC is less than the power dissipation of traditional CSVAC, a two-stage CSVAC must satisfy

$$P_{CSVAC}(A_{in}, G_1) + P_{CSVAC}(A_{in}G_1, G/G_1) < P_{CSVAC}(A_{in}, G). \tag{S42}$$

Based on (S31A) and (S41), there is $G > 1.156$. When the total voltage gain of multistage CSVACs exceeds 1.156 and $A_{in}$ is sufficiently small, i.e., satisfying $v \doteq 0$ when $c \gg b$, the multistage structure can reduce the power dissipation of CSVACs.

For the multistage CSVAC with the number of stages as $K$, the power dissipation is expressed as

$$P_{sum}^{(K)} = P_{CSVAC}(A_{in}, G_1) + P_{CSVAC}(A_{in}G_1, G_2) + \cdots + P_{CSVAC}(A_{in}\prod_{\lambda=1}^{K-1} G_\lambda, G \Big/ \prod_{\lambda=1}^{K-1} G_\lambda). \tag{S43}$$

For fixed $A_{in}$ and $G$, $P_{sum}^{(K)}$ is the function of $G_1, G_2, ..., G_{K-1}$, i.e., $P_{sum}^{(K)}(G_1, G_2, ..., G_{K-1})$. Since $P_{sum}^{(K)}$ is the nonnegative summation of $P_{CSVAC}(\cdot)$, $P_{sum}^{(K)}(G_1, G_2, ..., G_{K-1})$ is the convex function of $G_1, G_2, ..., G_{K-1}$[4]. To achieve the minimum power dissipation of a $K$-stage CSVAC with fixed $A_{in}$ and $G$, the necessary and sufficient condition is

$$\frac{\partial P_{sum}^{(K)}(G_1, G_2, ..., G_{K-1})}{\partial G_\lambda} = 0. \tag{S44}$$

where $\lambda \in \{1, 2, ..., (K-1)\}$. Assuming $A_{in}$ is small enough, i.e., satisfying $v \doteq 0$ when $c \gg b$, (S44) is equivalent to

$$G_1 = G_2 = \cdots = G_{K-1} = G \Big/ \prod_{\lambda=1}^{K-1} G_\kappa, \tag{S45}$$

i.e.,

$$G_\lambda = G^{1/K}. \tag{S46}$$

If $P_{\text{sum}}^{(K)}$ satisfies (S45) and

$$P_{\text{sum}}^{(\kappa)} < P_{\text{sum}}^{(\kappa-1)}, \tag{S47}$$

where $\kappa \in \{2, 3, ..., K\}$, then the power dissipation of multistage CSVACs with the number of stages as $K$ is less than the power dissipation of CSVACs with numbers of stages less than $K$. Based on (S31A), (S46), and (S47), there is $G_1 = G_2 = \cdots = G_{K-1} = G \Big/ \prod_{\lambda=1}^{K-1} G_\lambda > 1.075$, i.e., $G > 1.075^K$. When the total voltage gain of multistage CSVACs exceeds $1.075^K$ and $A_{\text{in}}$ is sufficiently small, i.e., satisfying $v \doteq 0$ when $c \gg b$, the power dissipation of multistage CSVACs with the number of stages as $K$ is less than the power dissipation of multistage CSVACs with numbers of stages less than $K$.

**Physical Entity Multistage CSVACs**

The power dissipation of physical CSVACs is calculated by the product of the difference between the positive and negative power supply voltages and the current between the positive and negative power supply, i.e.,

$$P_{\text{entity}} = \left[(+V_d) - (-V_d)\right] J_D = 2V_d J_D, \tag{S48}$$

where $J_D$ is the current between the positive and negative power supply. Based on the experimental data, the power dissipation of physical CSVACs could be expressed as a function of $A_{\text{in}}$ and $G$, i.e.,

$$P_{\text{entity}}(A_{\text{in}}, G) = \exp(a_{\text{entity}} + b_{\text{entity}} A_{\text{in}} + c_{\text{entity}} G), \tag{S49A}$$

where

$$\begin{cases} a_{\text{entity}} = -50.5 \\ b_{\text{entity}} = 1.415 \\ c_{\text{entity}} = 33.49 \end{cases}. \quad (S49B)$$

For a physical multistage CSVAC with the number of stages as $K$, the power dissipation is expressed as

$$P^{(K)}_{\text{sum\_entity}} = P_{\text{entity}}(A_{\text{in}}, G_1) + P_{\text{entity}}(A_{\text{in}} G_1, G_2) + \cdots + P_{\text{entity}}(A_{\text{in}} \prod_{\lambda=1}^{K-1} G_\lambda, G / \prod_{\lambda=1}^{K-1} G_\lambda). \quad (S50)$$

For fixed $A_{\text{in}}$ and $G$, $P^{(K)}_{\text{sum\_entity}}$ is the function of $G_1, G_2, ..., G_{K-1}$, i.e., $P^{(K)}_{\text{sum\_entity}}(G_1, G_2, ..., G_{K-1})$. To achieve the minimum power dissipation of a $K$-stage CSVAC with fixed $A_{\text{in}}$ and $G$, the necessary and sufficient condition is

$$\frac{\partial P^{(K)}_{\text{sum\_entity}}(G_1, G_2, ..., G_{K-1})}{\partial G_\lambda} = 0. \quad (S51)$$

The relationship between the minimum power dissipation and the number of stages of physical multistage CSVACs is shown in Figure S2. Based on the results in Figure S2, the optimal number of stages is 2 for the physical multistage CSVAC. Based on (S51), the optimal voltage gain at the first stage and the second stage are respectively 1.16 and 1.12, i.e., $G_1^{\text{opt}} = 1.16$ and $G_2^{\text{opt}} = 1.12$, when the optimal stage number is 2.

# SUPPLEMENTAL FIGURES AND LEGEND

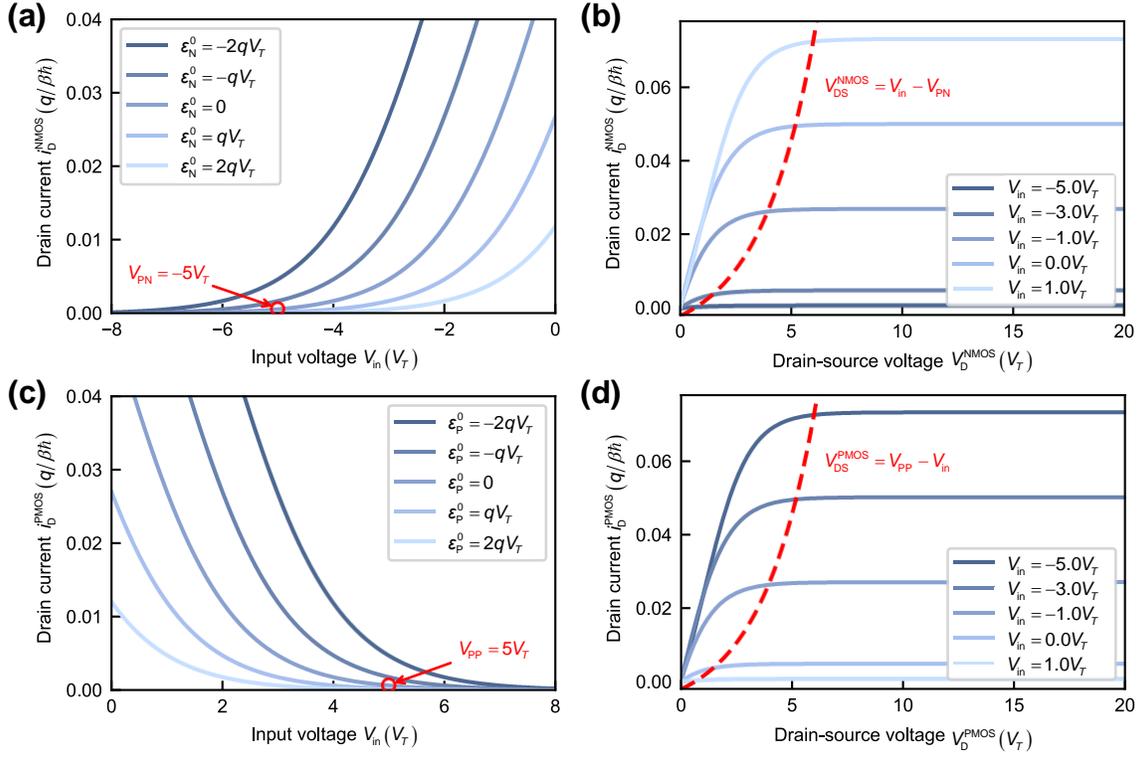

**Figure S1. Stochastic thermodynamic transistor current-voltage characteristics.** (**a**) Transfer characteristics of NMOS transistors. (**b**) Output characteristics of NMOS transistors. (**c**) Transfer characteristics of PMOS transistors. (**d**) Output characteristics of PMOS transistors.

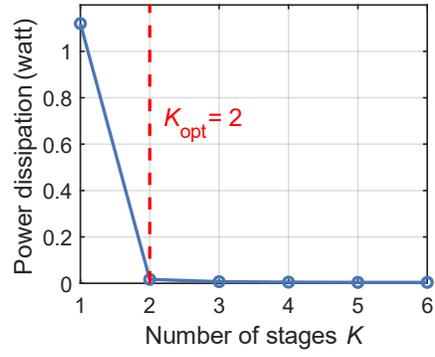

**Figure S2. Relationship between the minimum power dissipation and the number of stages of physical multistage CSVACs.**

**SUPPLEMENTAL REFERENCES**